\documentclass[12pt]{iopart}

\eqnobysec
\usepackage{graphicx}
\usepackage{times}

\usepackage{color}

\newcommand{\mean}[1]{\langle #1 \rangle}

\begin{document}

\title[2D arrays of superconducting and soft magnetic strips]{%
Two-dimensional arrays of superconducting and soft magnetic strips as dc magnetic metamaterials
}

\author{Yasunori Mawatari}
\address{%
National Institute of Advanced Industrial Science and Technology, \\
Tsukuba, Ibaraki 305-8568, Japan
}

\begin{abstract}
We have theoretically investigated the magnetic response of two-dimensional (2D) arrays of superconducting and soft magnetic strips, which are regarded as models of dc magnetic metamaterials. 
The anisotropy of the macroscopic permeabilities depends on whether the applied magnetic field is parallel to the wide surface of the strips ($\mu_{\parallel}$) or perpendicular ($\mu_{\perp}$). 
For the 2D arrays of superconducting strips, $0<\mu_{\perp}/\mu_0\ll \mu_{\parallel}/\mu_0\simeq 1$, whereas for the 2D arrays of soft magnetic strips, $\mu_{\parallel}/\mu_0\gg\mu_{\perp}/\mu_0\simeq 1$, where $\mu_0$ is the vacuum permeability.  
We also demonstrate that strong anisotropy of the macroscopic permeability can be obtained for hybrid arrays of superconducting and soft magnetic strips,  where $\mu_{\parallel}/\mu_0\gg 1\gg \mu_{\perp}/\mu_0>0$. 
\end{abstract}
\submitto{\SUST}

\section{Introduction}
It has been proposed that dc magnetic metamaterials can be used for magnetic field control~\cite{Wood07,Magnus08,Navau09}, and their application to magnetic cloaking devices has been investigated~\cite{Cummer06,Schurig06,Yaghjian08,Sanchez11,Narayana11,Gomory12}. 
Arrays of thin superconductors are candidates for dc magnetic metamaterials, because their magnetic permeability can exhibit geometrical anisotropy; the macroscopic permeability is small (i.e., $\mu_{\perp}/\mu_0\ll 1$) when the applied magnetic field is perpendicular to the wide surface of the thin superconductors, whereas thin superconductors are magnetically transparent (i.e., $\mu_{\parallel}/\mu_0\simeq 1$) when an applied field is parallel to the wide surface~\cite{Wood07,Magnus08,Navau09,Mawatari12}. 
The behavior of the arrays of thin soft magnets is analogously dual to that of the arrays of thin superconductors; thin soft magnets have large permeability (i.e., $\mu_{\parallel}/\mu_0\gg 1$) when the applied magnetic field is parallel to the wide surface of thin soft magnets, whereas thin soft magnets are magnetically transparent (i.e., $\mu_{\perp}/\mu_0\simeq 1$) when the applied field is perpendicular to the wide surface. 
Because of the anisotropy in the macroscopic permeability, arrays of thin superconductors and soft magnets can behave as dc magnetic metamaterials and can be used to control dc magnetic fields. 

In this paper we theoretically investigate the distribution of the magnetic field in two-dimensional (2D) arrays of superconducting strips and of soft magnetic strips, and present analytical expressions for the macroscopic permeabilities that characterize the magnetic response of the 2D arrays. 
We propose hybrid arrays of superconducting and soft magnetic strips that have both small perpendicular permeability, $0<\mu_{\perp}/\mu_0\ll 1$, and large parallel permeability, $\mu_{\parallel}/\mu_0\gg 1$. 
This paper is organized as follows: the basic formalism for the two-dimensional magnetic field is laid out in Sec.~2, the results for the 2D arrays of superconducting strips~\cite{Mawatari12} are shown in Sec.~3, the 2D arrays of soft magnetic strips are investigated in Sec.~4, the hybrid arrays of superconducting and soft magnetic strips are examined in Sec.~5, and a brief  discussion and summary of the results are given in Sec.~6.

\section{Two-dimensional magnetic field
\label{sec:2D-field}}
\subsection{Local (microscopic) magnetic field}
We investigate 2D arrays of superconducting and soft magnetic strips as the basic components of dc magnetic metamaterials. 
The thickness, $d$, of the strips is much smaller than the width, and is regarded as infinitesimal, $\epsilon=d/2\to 0$. 
The length, $L_z$, of the strips along the $z$ axis is much larger than the width, and is regarded as infinite, $L_z\to\infty$. 
The wide surface of the strips is parallel to the $xz$ plane, and the strips are regularly arranged in the $xy$ plane. 
We analyze the local (microscopic) magnetic field, ${\bi H}=H_x(x,y)\hat{\bi x} +H_y(x,y)\hat{\bi y}$, in the $xy$ plane. 
Outside the strips, the relationship between the local magnetic field, ${\bi H}$, and the local magnetic induction, ${\bi B}=B_x(x,y)\hat{\bi x} +B_y(x,y)\hat{\bi y}$, is given by ${\bi B}=\mu_0{\bi H}$, where $\mu_0$ is the vacuum permeability. 

The 2D magnetic field is analyzed using the complex field~\cite{Landau-Lifschitz,Beth66} 
\begin{equation}
	{\cal H}(\zeta) = H_y(x,y) +\rmi H_x(x,y), 
\label{eq:complex-H(z)}
\end{equation}
as the analytic function of the complex variable $\zeta=x+\rmi y$. 
The complex potential is defined by ${\cal G}(\zeta)=\int{\cal H}(\zeta)\rmd\zeta$, and the contour lines of $\mbox{Re}\,{\cal G}(x+\rmi y)$ correspond to the magnetic field lines in the $xy$ plane.

\subsection{Macroscopic field and macroscopic permeability}
In the unit cell of the 2D array, the macroscopic magnetic field $\mean{\bi H}$ is calculated as the averaged line integral of $\bi H$ at the cell edge, whereas the macroscopic magnetic permeability $\mean{\bi B}$ is calculated as the averaged surface integral of $\bi B$ at the cell side~\cite{Mawatari12,Pendry99,Smith06}. 
Because of the different definitions of the averaging procedure for obtaining the macroscopic fields, the macroscopic relationship, $\mean{\bi B}\neq\mu_0\mean{\bi H}$, generally holds, even though the microscopic relationship, ${\bi B}=\mu_0{\bi H}$, holds. 
We consider the case where the wide surfaces of the strips are parallel to the $xz$ plane; therefore the permeability tensor $\mu_{\alpha\beta}$ defined by $\mean{B_{\alpha}}=\mu_{\alpha\beta}\mean{H_{\beta}}$ has only diagonal components, $\mu_{xx}=\mu_{\parallel}$ and $\mu_{yy}=\mu_{\perp}$: 
\begin{equation}
	\mean{B_x}= \mu_{\parallel}\mean{H_x} \quad\mbox{and}\quad
	\mean{B_y}= \mu_{\perp}\mean{H_y} . 
\label{eq:B-H-mu}
\end{equation}
The magnetic response to a parallel field is characterized by the parallel permeability, $\mu_{\parallel}$, whereas the response to a perpendicular field is characterized by the perpendicular permeability, $\mu_{\perp}$. 
We demonstrate later that for superconducting strip arrays, $0<\mu_{\perp}/\mu_0\ll \mu_{\parallel}/\mu_0 \simeq 1$, whereas for soft magnetic strip arrays, $\mu_{\parallel}/\mu_0\gg \mu_{\perp}/\mu_0 \simeq 1$. 
We also show that for the hybrid arrays of superconducting and soft magnetic strips, $0<\mu_{\perp}/\mu_0\ll 1\ll \mu_{\parallel}/\mu_0$.

\section{Two-dimensional arrays of superconducting strips
\label{sec:SC-array}}
In this section we briefly review the magnetic field distribution and macroscopic permeability of 2D arrays of superconducting strips reported in Ref.~\cite{Mawatari12}. 
Each superconducting strip has a width of $2w$, an infinitesimal thickness of $d$ (i.e., $\epsilon=d/2\to 0$), and an infinite length along the $z$ axis. 
The wide surfaces of the superconducting strips are parallel to the $xz$ plane. 
It is assumed that the superconducting strips are in the complete shielding state, where the magnetic field is completely shielded in the superconducting strips. 
The complete shielding state is achieved when the London penetration depth, $\lambda$, is much smaller than the dimensions of the superconducting strips, $\lambda/d\to 0$ for thick strips or $\lambda^2/wd\to 0$ for thin strips, in the Meissner state. 
The complete shielding state has also been observed for a weak field or large critical current density limit in the critical state model ~\cite{Bean62}. 
The 2D arrays of superconducting strips are exposed to an applied magnetic field of ${\bi H}_a=H_{ax}\hat{\bi x} +H_{ay}\hat{\bi y}$, which is expressed as $H_{ay}+\rmi H_{ax}$ in terms of the complex field. 

When a 2D array of superconducting strips is exposed to a {\em parallel} magnetic field along the $x$ axis, the magnetic field is not disturbed by thin superconducting strips for which $\epsilon\to 0$. 
Therefore, the macroscopic permeability for a parallel field is equal to the vacuum permeability, $\mu_{\parallel}/\mu_0=1$, for the thin-strip limit.  

In contrast, when a 2D array of superconducting strips is exposed to a {\em perpendicular} magnetic field along the $y$ axis, the magnetic field is disturbed by the superconducting strips. 
Because of the magnetic shielding by the superconducting strips, the macroscopic permeability for a perpendicular field is smaller than the vacuum permeability, $0<\mu_{\perp}/\mu_0<1$, depending on the geometry of the 2D array.

\subsection{Rectangular array of superconducting strips}
\begin{figure}[bthp]
\center\includegraphics{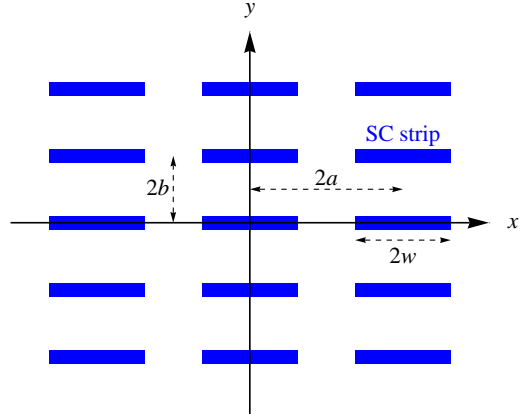}
\caption{%
Rectangular array of superconducting (SC) strips. 
The solid horizontal bars show the cross section of the superconducting strips in the $xy$ plane. 
In the $n$th layer at $y=2nb$, the $m$th strip is situated at $|x-2ma|<w$, where $m=0,\,\pm 1,\,\pm2,\dots,\,\pm\infty$, and $n=0,\,\pm 1,\,\pm2,\dots,\,\pm\infty$. 
}\label{fig:SC-rectangular}
\end{figure}

We consider a rectangular array of superconducting strips, in which the superconducting strips of width $2w$ are regularly arranged with a unit cell of $2a\times 2b$ in the $xy$ plane, as shown in \fref{fig:SC-rectangular}. 

We employ the auxiliary complex variable, $\eta_r$, defined as 
\begin{equation}
	\eta_r(\zeta)\equiv \mbox{sn}(\zeta/c_r,k_r), 
\label{eq:eta-rectangular}
\end{equation}
where $\mbox{sn}(u,k)$ is the sine amplitude (i.e., the Jacobi sn function) \cite{Gradshtein-Ryzhik}. 
The modulus, $k_r$, is obtained as a function of $b/a$ by solving 
\begin{equation}
	b/a = {\bf K}(\sqrt{1-k_r^2})/{\bf K}(k_r), 
\label{eq:b/a-K_rectangular}
\end{equation}
where ${\bf K}(k)$ is the complete elliptic integral of the first kind~\cite{Gradshtein-Ryzhik}. 
The $c_r$ in \eref{eq:eta-rectangular} is then given by 
\begin{equation}
	c_r = a/{\bf K}(k_r) = b/{\bf K}(\sqrt{1-k_r^2}) . 
\label{eq:c-rectangular}
\end{equation}
The complex field ${\cal H}(\zeta)$ and the complex potential ${\cal G}(\zeta)=\int_{ib}^{\zeta}{\cal H}(\zeta')\rmd\zeta'$ for the rectangular array of superconducting strips in the complete shielding state are ~\cite{Mawatari12} 
\begin{eqnarray}
	{\cal H}(\zeta) &= H_{0y}\frac{\eta_r(\zeta)}{\sqrt{\eta_r(\zeta)^2-\gamma_r^2}} 
		+\rmi H_{0x} , 
\label{eq:H(z)_rectangular}\\
	{\cal G}(\zeta) &= \frac{H_{0y}c_r}{\sqrt{1-k_r^2\gamma_r^2}}\, 
		F\left(\arcsin\sqrt{\frac{k_r^{-2}-\gamma_r^2}{\eta_r(\zeta)^2-\gamma_r^2}}, 
		\kappa_r \right) 
		+\rmi H_{0x}\zeta , 
\label{eq:G(z)_rectangular}
\end{eqnarray}
where $F(\varphi,k)$ is the elliptic integral of the first kind~\cite{Gradshtein-Ryzhik}. 
The parameters $\gamma_r$ and $\kappa_r$ in \eref{eq:H(z)_rectangular} and \eref{eq:G(z)_rectangular} are defined as 
\begin{eqnarray}
	\gamma_r &= \eta_r(w)= \mbox{sn}(w/c_r,k_r) , 
\label{eq:gamma_r}\\
	\kappa_r &= \sqrt{\frac{1-\gamma_r^2}{k_r^{-2}-\gamma_r^2}} 
		= \frac{k_r \mbox{cn}(w/c_r,k_r)}{\mbox{dn}(w/c_r,k_r)} , 
\label{eq:kappa_r}
\end{eqnarray}
where $\mbox{cn}(u,k)$ and $\mbox{dn}(u,k)$ are the Jacobi cn and dn functions, respectively. 
We do not need to consider the details of the constants $H_{0y}$ and $H_{0x}$ in \eref{eq:H(z)_rectangular} and \eref{eq:G(z)_rectangular}, because neither $H_{0x}$ nor $H_{0y}$ affects the final results of the effective permeability~\cite{H0-Ha}. 

When the rectangular array of superconducting strips is exposed to a {\em parallel} magnetic field along the $x$ axis, the magnetic field is not disturbed by thin superconducting strips where $\epsilon\to 0$; that is, \eref{eq:H(z)_rectangular} shows that ${\cal H}(\zeta)=\rmi H_{0x}$ for $H_{0y}=0\neq H_{0x}$. 
In this case, the macroscopic fields are $\mean{B_x}/\mu_0=\mean{H_x}=H_{0x}$, and the macroscopic permeability for a parallel field is equal to the vacuum permeability, $\mu_{\parallel}/\mu_0=1$, for the thin-strip limit.  

\begin{figure}[bthp]
\center\includegraphics{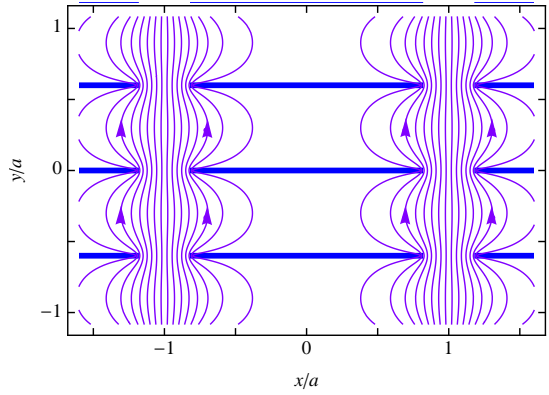}
\caption{%
Magnetic field lines in the rectangular array of superconducting strips (shown as solid horizontal bars) exposed to a perpendicular magnetic field for $w/a=0.8$ and $b/a=0.3$. 
}\label{fig:field-lines_SC-rect}
\end{figure}

In contrast, when the rectangular array of superconducting strips is exposed to a {\em perpendicular} magnetic field along the $y$ axis (i.e., $H_{0x}=0\neq H_{0y}$), the magnetic field is disturbed by the superconducting strips. 
Because of the magnetic shielding by the superconducting strips, the macroscopic permeability for a perpendicular field is smaller than the vacuum permeability, $0<\mu_{\perp}/\mu_0<1$, depending on the geometry of the 2D array. 
\Fref{fig:field-lines_SC-rect} shows the magnetic field lines as the contour lines of $\mbox{Re}\,{\cal G}(x+\rmi y)$ obtained from \eref{eq:G(z)_rectangular} for $H_{0x}=0$. 
The magnetic field is concentrated near the gaps between the edges of the superconducting strips. 

The local magnetic induction, $B_y(x,y)=\mu_0 H_y(x,y)$, and the local magnetic field, $H_y(x,y)=\mbox{Re}\,{\cal H}(x+\rmi y)$, are obtained from \eref{eq:H(z)_rectangular}. 
We examine the macroscopic perpendicular fields, $\mean{B_y}$ and $\mean{H_y}$, averaged over the unit cell of the rectangular array. 
The macroscopic magnetic induction $\mean{B_y}$ and macroscopic magnetic field $\mean{H_y}$ are calculated from the local fields as~\cite{Mawatari12} 
\begin{eqnarray}
	\frac{\mean{B_y}}{\mu_0} &\equiv  \frac{1}{2a}\int_{-a}^{+a} H_y(x,b)\rmd x 
	= \frac{1}{2a}\int_{-a}^{+a} H_y(x,y)\rmd x , 
\label{eq:<By>_rectangular}\\
	\mean{H_y} &\equiv  
		\frac{1}{2b}\int_{0}^{2b} H_y(a,y)\rmd y . 
\label{eq:<Hy>_rectangular}
\end{eqnarray}
The last expression of \eref{eq:<By>_rectangular} is independent of $y$, because $\nabla\cdot{\bi B}=0$~\cite{Mawatari12}. 
As shown in \ref{app:sc-rect}, the macroscopic fields defined by \eref{eq:<By>_rectangular} and \eref{eq:<Hy>_rectangular} are consistent with the macroscopic relationship, 
\begin{equation}
	\mean{B_y}/\mu_0 = \mean{H_y} +\mean{M_y} , 
\label{eq:By-Hy-My}
\end{equation}
where $\mean{M_y}$ is the magnetization of superconducting strips defined by 
\begin{equation}
	\mean{M_y} \equiv -\frac{1}{4ab}\int_{-w}^{+w} xK_z(x)\rmd x 
\label{eq:My}
\end{equation}
and $K_z(x)=H_x(x,-\epsilon)-H_x(x,+\epsilon)$ is the sheet current density in superconducting strips. 

\begin{figure}[bthp]
\center\includegraphics{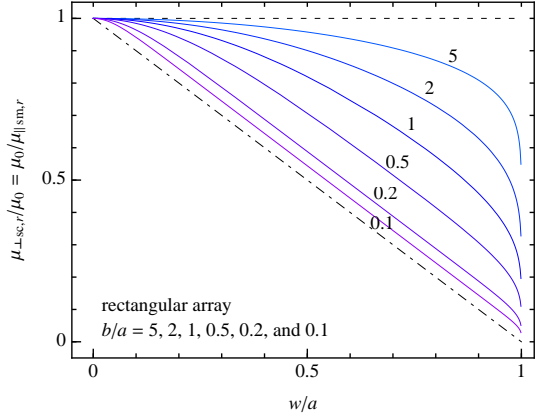}
\caption{%
Effective permeability of the rectangular array of superconducting strips in a perpendicular field, $\mu_{\perp\rm sc,r}$, as a function of $w/a$ for $b/a=5,\,2,\,1,\,0.5,\,0.2$, and $0.1$. 
The dashed line corresponds to $\mu_{\perp\rm sc,r}/\mu_0=1$ for $b/a\to \infty$ and the chained line corresponds to $\mu_{\perp\rm sc,r}/\mu_0=1-w/a$ for $b/a\to 0$. 
The effective permeability of the rectangular array of soft magnetic strips in a parallel magnetic field, $\mu_{\parallel\rm sm,r}$, corresponds to the inverse of $\mu_{\perp\rm sc,r}$; that is, $\mu_{\perp\rm sc,r}/\mu_0=\mu_{\parallel\rm sm,r}/\mu_0$. 
}\label{fig:mu-perp_SC-rect}
\end{figure}

The macroscopic perpendicular permeability $\mu_{\rm \perp sc,r}= \mean{B_y}/\mean{H_y}$ for the rectangular array of superconducting strips is obtained from \eref{eq:H(z)_rectangular}, \eref{eq:<By>_rectangular}, and \eref{eq:<Hy>_rectangular}, as 
\begin{equation}
	\frac{\mu_{\perp\rm sc,r}}{\mu_0}= 
		\frac{b}{a}\frac{{\bf K}(\kappa_r)}{{\bf K}(\sqrt{1-\kappa_r^2})} , 
\label{eq:mu-perp_SC-rect}
\end{equation}
where $\kappa_r$ is given by \eref{eq:kappa_r}. 
Simple expressions of $\mu_{\perp\rm sc,r}$ for limiting cases can be obtained from \eref{eq:mu-perp_SC-rect}.
For a large stack spacings, $b/a>2$,
\begin{equation}
	\frac{\mu_{\perp\rm sc,r}}{\mu_0} 
	\simeq \left[1-\frac{2a}{\pi b}\ln\cos\left(\frac{\pi w}{2a}\right)\right]^{-1} , 
\label{eq:mu-SC_large-spacing}
\end{equation}
whereas for small stack spacings, $b/a\ll 1$,
\begin{equation}
	\frac{\mu_{\perp\rm sc,r}}{\mu_0} 
	\simeq 1-\frac{w}{a}+\frac{2b}{\pi a}\ln 2 . 
\label{eq:mu-SC-rect_small-spacing}
\end{equation}
\Eref{eq:mu-SC-rect_small-spacing} is not accurate near $w/a\simeq 0$ or $1$. 
\Fref{fig:mu-perp_SC-rect} shows plots of $\mu_{\perp\rm sc,r}/\mu_0$ versus $w/a$ obtained from \eref{eq:b/a-K_rectangular}, \eref{eq:c-rectangular}, \eref{eq:kappa_r}, and \eref{eq:mu-perp_SC-rect}. 
We can obtain a small perpendicular permeability, $\mu_{\perp\rm sc,r}/\mu_0\ll 1$, when the gaps between the edges of the superconducting strips are small, $1-w/a\ll 1$. 

\subsection{Hexagonal array of superconducting strips}
\begin{figure}[bthp]
\center\includegraphics{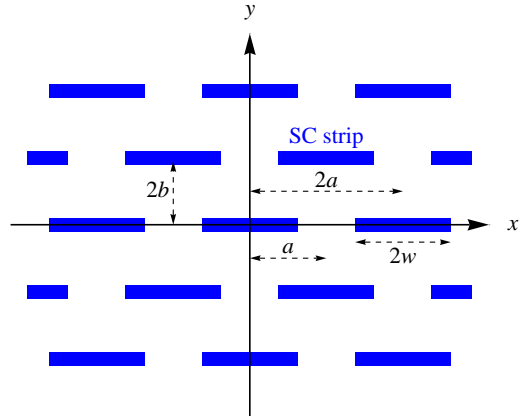}
\caption{%
Hexagonal array of superconducting strips. 
The solid horizontal bars show the cross section of superconducting strips in the $xy$ plane. 
In the even layer at $y=4nb$, the $m$th strip is situated at $|x-2ma|<w$, whereas in the odd layer at $y=(4n+2)b$, the $m$th strip is situated at $|x-(2m+1)a|<w$, where $m=0,\,\pm 1,\,\pm2,\dots,\,\pm\infty$ and $n=0,\,\pm 1,\,\pm2,\dots,\,\pm\infty$. 
}\label{fig:SC-hexagonal}
\end{figure}

We next consider a hexagonal array of superconducting strips, in which the superconducting strips of width $2w$ are regularly arranged in the $xy$ plane, as shown in \fref{fig:SC-hexagonal}. 

We employ the auxiliary complex variable, $\eta_h$, defined as 
\begin{equation}
	\eta_h(\zeta)\equiv \mbox{sn}(\zeta/c_h,k_h) .
\label{eq:eta-hexagonal}
\end{equation}
The modulus, $k_h$, is obtained as a function of $b/a$ by solving 
\begin{equation}
	2b/a = {\bf K}(\sqrt{1-k_h^2})/{\bf K}(k_h) .
\label{eq:b/a-K_hexagonal}
\end{equation}
The relationship between $k_h$ defined by \eref{eq:b/a-K_hexagonal} and $k_r$ defined by \eref{eq:b/a-K_rectangular} is expressed by $k_h=(1-\sqrt{1-k_r^2})/(1+\sqrt{1-k_r^2})$. 
The value of $c_h$ in \eref{eq:eta-hexagonal} is given by
\begin{equation}
	c_h = a/{\bf K}(k_h) = 2b/{\bf K}(\sqrt{1-k_h^2}) . 
\label{eq:c-hexagonal}
\end{equation}
The complex field ${\cal H}(\zeta)$ and the complex potential ${\cal G}(\zeta)=\int_{ib}^{\zeta}{\cal H}(\zeta')\rmd\zeta'$ for the rectangular array of superconducting strips in the complete shielding state are~\cite{Mawatari12} 
\begin{eqnarray}
	{\cal H}(\zeta) &= H_{0y}\frac{\eta_h(\zeta)\sqrt{\eta_h(\zeta)^2-k_h^{-2}}}{%
		\sqrt{\eta_h(\zeta)^2-\gamma_h^2}\sqrt{\eta_h(\zeta)^2-\beta_h^2}} 
		+\rmi H_{0x} , 
\label{eq:H(z)_hexagonal}\\
	{\cal G}(\zeta) &= \frac{H_{0y}c_h}{k_h\sqrt{\beta_h^2-\gamma_h^2}}\, 
		F\left(\arcsin\sqrt{\frac{\beta_h^2-\gamma_h^2}{\eta_h(\zeta)^2-\gamma_h^2}}, 
		\kappa_h \right) 
		+\rmi H_{0x}\zeta , 
\label{eq:G(z)_hexagonal}
\end{eqnarray}
where 
\begin{eqnarray}
	\gamma_h &= \eta_h(w)= \mbox{sn}(w/c_h,k_h) , 
\label{eq:gamma_h}\\
	\beta_h &= \eta_h(a-w+2\rmi b) 
	= \sqrt{\frac{k_h^{-2}-\gamma_h^2}{1-\gamma_h^2}} 
		= \frac{\mbox{dn}(w/c_h,k_h)}{k_h \mbox{cn}(w/c_h,k_h)} . 
\label{eq:beta_h}\\
	\kappa_h &= \left[\frac{(1-\gamma_h^2)^2}{k_h^{-2}-1+(1-\gamma_h^2)^2}\right]^{1/2} 
		= \left[1+\frac{k_h^{-2}-1}{\mbox{cn}^4(w/c_h,k_h)}\right]^{-1/2} . 
\label{eq:kappa_h}
\end{eqnarray}

Under a {\em parallel} magnetic field along the $x$ axis, \eref{eq:H(z)_hexagonal} shows that ${\cal H}(\zeta)=\rmi H_{0x}$ for $H_{0y}=0\neq H_{0x}$, and the macroscopic permeability for a parallel field is equal to the vacuum permeability, $\mu_{\parallel}/\mu_0=1$, for the thin-strip limit.  

\begin{figure}[bthp]
\center\includegraphics{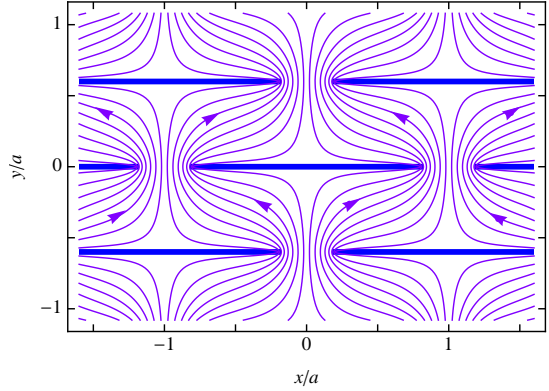}
\caption{%
Magnetic field lines in the hexagonal arrays of superconducting strips (shown as solid horizontal bars) exposed to a perpendicular magnetic field for $w/a=0.8$ and $b/a=0.3$. 
}\label{fig:field-lines_SC-hexa}
\end{figure}

In contrast, under a {\em perpendicular} magnetic field along the $y$ axis (i.e., $H_{0x}=0\neq H_{0y}$), the magnetic field is disturbed by the superconducting strips. 
\Fref{fig:field-lines_SC-hexa} shows the magnetic field lines as the contour lines of $\mbox{Re}\,{\cal G}(x+\rmi y)$ obtained from \eref{eq:G(z)_hexagonal} for $H_{0x}=0$. 
The magnetic field  is concentrated near the gaps between the edges of the superconducting strips. 

The definitions of the macroscopic magnetic induction $\mean{B_y}$ and the magnetization $\mean{M_y}$ for the hexagonal array are the same as those for the rectangular array, and are expressed by \eref{eq:<By>_rectangular} and \eref{eq:My}, respectively. 
The definition of the macroscopic magnetic field for the hexagonal array, $\mean{H_y}$, given by \eref{eq:<Hy>_rectangular} is inconsistent with the macroscopic relationship given by \eref{eq:By-Hy-My}. 
Therefore, we use the modified definition of $\mean{H_y}$ for the hexagonal array~\cite{Mawatari12}, 
\begin{equation}
	\mean{H_y}\equiv \frac{1}{2b}\left[\int_0^{2b}H_y(a,y)\rmd y 
		-\int_0^a H_x(x,2b-\epsilon)\rmd x\right] . 
\label{eq:<Hy>_hexagonal}
\end{equation}
For the hexagonal array, the macroscopic quantities defined by \eref{eq:<By>_rectangular}, \eref{eq:My}, and \eref{eq:<Hy>_hexagonal} satisfy \eref{eq:By-Hy-My}, as shown in \ref{app:sc-hexa}. 

\begin{figure}[bthp]
\center\includegraphics{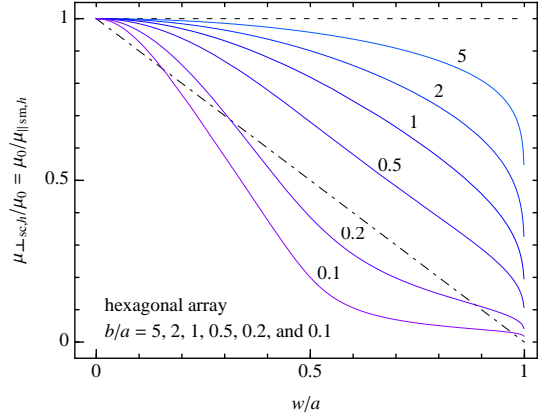}
\caption{%
Effective permeability of the hexagonal array of superconducting strips in a perpendicular field, $\mu_{\perp\rm sc,h}$, as a function of $w/a$ for $b/a=5,\,2,\,1,\,0.5,\,0.2$, and $0.1$. 
The dashed line corresponds to $\mu_{\perp\rm sc,h}/\mu_0=1$ for $b/a\to \infty$, and the chained line of $\mu_{\perp\rm sc,h}/\mu_0=1-w/a$ is shown for comparison with \fref{fig:mu-perp_SC-rect}. 
The effective permeability of the hexagonal array of soft magnetic strips in a parallel magnetic field, $\mu_{\parallel\rm sm,h}$, corresponds to the inverse of $\mu_{\perp\rm sc,h}$; that is, $\mu_{\perp\rm sc,h}/\mu_0=\mu_0/\mu_{\parallel\rm sm,h}$. 
}\label{fig:mu-perp_SC-hexa}
\end{figure}

The macroscopic perpendicular permeability, $\mu_{\rm \perp sc,h}= \mean{B_y}/\mean{H_y}$, for the hexagonal array of superconducting strips, is obtained from \eref{eq:<By>_rectangular}, \eref{eq:H(z)_hexagonal}, and \eref{eq:<Hy>_hexagonal}:
\begin{equation}
	\frac{\mu_{\perp\rm sc,h}}{\mu_0}= 
		\frac{2b}{a}\frac{{\bf K}(\kappa_h)}{{\bf K}(\sqrt{1-\kappa_h^2})} .
\label{eq:mu-perp_SC-hexa}
\end{equation}
Here $\kappa_h$ is given by \eref{eq:kappa_h}. 
For large stack spacings, $b/a>2$, the right-hand side of \eref{eq:mu-perp_SC-hexa} also reduces to the right-hand side of \eref{eq:mu-SC_large-spacing}. 
For small stack spacings, $b/a\ll 1$, 
\begin{equation}
	\frac{\mu_{\perp\rm sc,h}}{\mu_0}\simeq \cases{%
		1-\frac{2w}{a} +\frac{8b}{\pi a}\ln2 
		& for $\displaystyle 0<w/a<1/2$ \\
		\left(\frac{2b}{a}\right)^2 \left(\frac{2w}{a}-1 +\frac{8b}{\pi a}\ln2\right)^{-1} 
		& for $1/2<w/a<1$ } . 
\label{eq:mu-SC-hexa_small-spacing}
\end{equation}
\Eref{eq:mu-SC-hexa_small-spacing} is not accurate near $w/a\simeq 0$, $1/2$ or $1$. 
\Fref{fig:mu-perp_SC-hexa} shows plots of $\mu_{\perp\rm sc,h}/\mu_0$ versus $w/a$ obtained from \eref{eq:b/a-K_hexagonal}, \eref{eq:c-hexagonal}, \eref{eq:kappa_h}, and \eref{eq:mu-perp_SC-hexa}. 
We can obtain a small perpendicular permeability, $\mu_{\perp\rm sc,h}/\mu_0\ll 1$, for a wide range of $0.5<w/a<1$, when $b/a\ll 1$. 

\section{Two-dimensional arrays of soft magnetic strips
\label{sec:SM-array}}
We investigate the magnetic field distribution and macroscopic permeability of 2D arrays of soft magnetic strips. 
The dimensions of the soft magnetic strips are the same as those of the superconducting strips shown in Sec.~\ref{sec:SC-array}: 
each soft magnetic strip has a width of $2w$, an infinitesimal thickness of $d$ (i.e., $\epsilon=d/2\to 0$), and an infinite length along the $z$ axis. 
The wide surfaces of the soft magnetic strips are parallel to the $xz$ plane. 
The soft magnetic strips are treated as ideal soft magnets, with an infinite permeability, zero hysteresis, and an infinite saturation field~\cite{Mawatari08}. 
In the ideal soft magnet, the relationship between $\bi B$ and $\bi H$ is given by ${\bi B}=\mu_m{\bi H}$, where $\mu_m/\mu_0\to\infty$. 
Outside the ideal soft magnet, ${\bi H}={\bi B}/\mu_0$ has only a perpendicular component at the surface~\cite{Jackson}. 
The 2D arrays of soft magnetic strips are exposed to an applied magnetic field, ${\bi H}_a=H_{ax}\hat{\bi x} +H_{ay}\hat{\bi y}$, that is expressed in terms of the complex field as $H_{ay}+\rmi H_{ax}$.

When the 2D array of soft magnetic strips is exposed to a {\em perpendicular} magnetic field along the $y$ axis, the magnetic field is not disturbed by thin soft magnetic strips of $\epsilon\to 0$. 
Therefore, the macroscopic permeability for a perpendicular field is equal to the vacuum permeability, $\mu_{\perp}/\mu_0=1$, for the thin-strip limit.  

When the 2D array of soft magnetic strips is exposed to a {\em parallel} magnetic field along the $x$ axis, on the other hand, the magnetic field is disturbed by soft magnetic strips. 
The macroscopic permeability of a perpendicular field is larger than the vacuum permeability, $\mu_{\parallel}/\mu_0>1$, depending on the geometry of the 2D array.

\subsection{Rectangular array of soft magnetic strips}
\begin{figure}[bthp]
\center\includegraphics{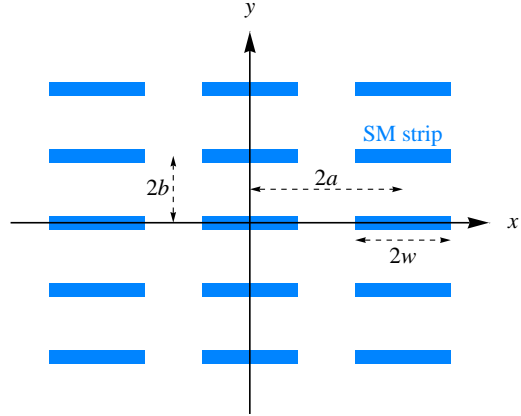}
\caption{%
Rectangular array of soft magnetic (SM) strips. 
The solid horizontal bars show the cross section of the soft magnetic strips in the $xy$ plane. 
In the $n$th layer at $y=2nb$, the $m$th strip is situated at $|x-2ma|<w$, where $m=0,\,\pm 1,\,\pm2,\dots,\,\pm\infty$ and $n=0,\,\pm 1,\,\pm2,\dots,\,\pm\infty$. 
}\label{fig:SM-rectangular}
\end{figure}

We consider a rectangular array of soft magnetic strips, in which soft magnetic strips of width $2w$ are regularly arranged with a unit cell of $2a\times 2b$ in the $xy$ plane, as shown in \fref{fig:SM-rectangular}. 
The geometry of the rectangular array of soft magnetic strips is exactly the same as that of the rectangular array of superconducting strips shown in \fref{fig:SC-rectangular}. 

The complex field, ${\cal H}(\zeta)$, and the complex potential, ${\cal G}(\zeta)=\int_{ib}^{\zeta}{\cal H}(\zeta')\rmd\zeta'$, for the rectangular array of soft magnetic strips based on the ideal soft magnet model are given by 
\begin{eqnarray}
	{\cal H}(\zeta) &= H_{0y} 
		+\rmi\,H_{0x}\frac{\eta_r(\zeta)}{\sqrt{\eta_r(\zeta)^2-\gamma_r^2}} , 
\label{eq:H(z)_rectangular_SM}\\
	{\cal G}(\zeta) &=  H_{0y}\zeta 
		+\frac{\rmi\,H_{0x}c_r}{\sqrt{1-k_r^2\gamma_r^2}}\, 
		F\left(\arcsin\sqrt{\frac{k_r^{-2}-\gamma_r^2}{\eta_r(\zeta)^2-\gamma_r^2}},  
		\kappa_r \right) , 
\label{eq:G(z)_rectangular_SM}
\end{eqnarray}
where $\eta_r$, $k_r$, $c_r$, $\gamma_r$, and $\kappa_r$ are defined by \eref{eq:eta-rectangular}, \eref{eq:b/a-K_rectangular}, \eref{eq:c-rectangular}, \eref{eq:gamma_r}, and \eref{eq:kappa_r}, respectively. 
The behavior of the soft magnetic strips is analogously dual to that of the superconducting strips; \eref{eq:H(z)_rectangular_SM} and \eref{eq:G(z)_rectangular_SM} are obtained simply by exchanging $H_{0y}\leftrightarrow\rmi H_{0x}$ in \eref{eq:H(z)_rectangular} and \eref{eq:G(z)_rectangular}, respectively~\cite{Chen02}. 

When the rectangular array of soft magnetic strips is exposed to a {\em perpendicular} magnetic field along the $y$ axis, the magnetic field is not disturbed by thin soft magnetic strips where $\epsilon\to 0$; that is, \eref{eq:H(z)_rectangular_SM} shows that ${\cal H}(\zeta)=H_{0y}$ for $H_{0x}=0\neq H_{0y}$. 
In this case, the macroscopic fields are $\mean{B_y}/\mu_0=\mean{H_y}=H_{0y}$, and the macroscopic permeability for a perpendicular field is equal to the vacuum permeability, $\mu_{\perp}/\mu_0=1$, for the thin-strip limit. 

\begin{figure}[bthp]
\center\includegraphics{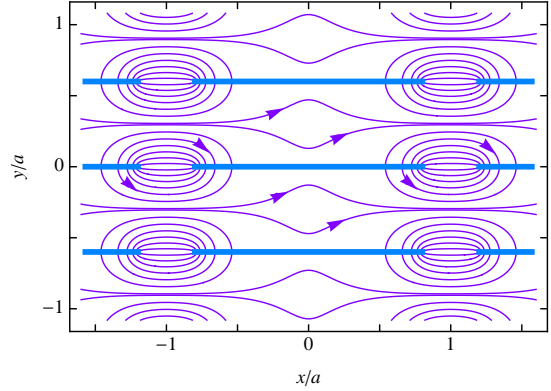}
\caption{%
Magnetic field lines in the rectangular array of soft magnetic strips (shown as solid horizontal bars) exposed to a parallel magnetic field for $w/a=0.8$ and $b/a=0.3$. 
}\label{fig:field-lines_SM-rect}
\end{figure}

In contrast, when the rectangular array of soft magnetic strips is exposed to a {\em parallel} magnetic field along the $x$ axis (i.e., $H_{0y}=0\neq H_{0x}$), the magnetic field is disturbed by the soft magnetic strips. 
The macroscopic permeability for a parallel field is larger than the vacuum permeability, $\mu_{\parallel}/\mu_0>1$, depending on the geometry of the 2D array. 
\Fref{fig:field-lines_SM-rect} shows the magnetic field lines as the contour lines of $\mbox{Re}\,{\cal G}(x+\rmi y)$ obtained from \eref{eq:G(z)_rectangular_SM} for $H_{0y}=0$. 

The macroscopic parallel fields, $\mean{B_x}$ and $\mean{H_x}$, averaged over the unit cell of the rectangular array are defined by 
\begin{eqnarray}
	\frac{\mean{B_x}}{\mu_0} &\equiv 
		\frac{1}{2b}\int_{0}^{2b} H_x(a,y)\rmd y , 
\label{eq:<Bx>_rectangular}\\
	\mean{H_x} &\equiv  \frac{1}{2a}\int_{-a}^{+a} H_x(x,b)\rmd x 
	= \frac{1}{2a}\int_{-a}^{+a} H_x(x,y)\rmd x . 
\label{eq:<Hx>_rectangular}
\end{eqnarray}
The last expression of \eref{eq:<Hx>_rectangular} is independent of $y$, because $\nabla\times{\bi H}=0$~\cite{Mawatari12}. 
As shown in \ref{app:sm-rect}, \eref{eq:<Bx>_rectangular} and \eref{eq:<Hx>_rectangular} are consistent with 
\begin{equation}
	\mean{B_x}/\mu_0 = \mean{H_x} +\mean{M_x} , 
\label{eq:Bx-Hx-Mx}
\end{equation}
where $\mean{M_x}$ is the magnetization arising from the soft magnetic strips, defined as~\cite{Mawatari12} 
\begin{equation}
	\mean{M_x}\equiv \frac{1}{4ab}\int_{-w}^{+w} x\sigma_m(x)\rmd x . 
\label{eq:Mx}
\end{equation}
The expression $\sigma_m(x)=H_y(x,+\epsilon)-H_y(x,-\epsilon)$ corresponds to the effective sheet magnetic charge in the soft magnetic strips~\cite{Mawatari08,Jackson}. 

The macroscopic parallel permeability, $\mu_{\rm \parallel sm,r}= \mean{B_x}/\mean{H_x}$, for the rectangular array of soft magnetic strips is obtained from \eref{eq:H(z)_rectangular_SM}, \eref{eq:<Bx>_rectangular}, and \eref{eq:<Hx>_rectangular}, as 
\begin{equation}
	\frac{\mu_{\parallel\rm sm,r}}{\mu_0}= 
		\frac{a}{b}\frac{{\bf K}(\sqrt{1-\kappa_r^2})}{{\bf K}(\kappa_r)} , 
\label{eq:mu-para_SM-rect}
\end{equation}
where $\kappa_r$ is given by \eref{eq:kappa_r}. 
Note that $\mu_{\perp\rm sc,r}$ given by \eref{eq:mu-perp_SC-rect} and $\mu_{\parallel\rm sm,r}$ given by \eref{eq:mu-para_SM-rect} hold the simple relationship $\mu_{\parallel\rm sm,r}=\mu_0^2/\mu_{\perp\rm sc,r}$. 
\Fref{fig:mu-perp_SC-rect} shows plots of $\mu_0/\mu_{\parallel\rm sm,r}$ versus $w/a$ obtained from \eref{eq:b/a-K_rectangular}, \eref{eq:c-rectangular}, \eref{eq:kappa_r}, and \eref{eq:mu-para_SM-rect}. 
We can obtain a large parallel permeability, $\mu_{\parallel\rm sm,r}/\mu_0\gg 1$, when the gaps between the edges of the soft magnetic strips are small, $1-w/a\ll 1$. 

\subsection{Hexagonal array of soft magnetic strips}
\begin{figure}[bthp]
\center\includegraphics{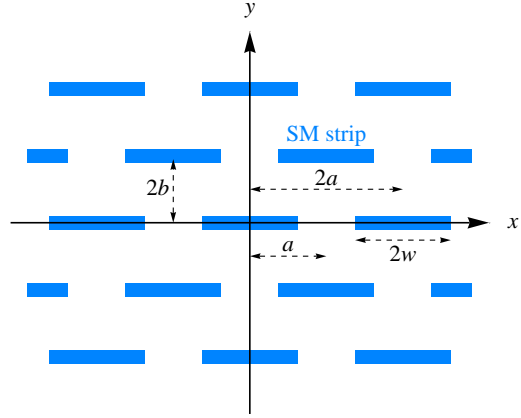}
\caption{%
Hexagonal array of soft magnetic strips. 
Solid horizontal bars show the cross sections of the soft magnetic strips in the $xy$ plane. 
In the even layer at $y=4nb$, the $m$th strip is situated at $|x-2ma|<w$, whereas in the odd layer at $y=(4n+2)b$, the $m$th strip is situated at $|x-(2m+1)a|<w$, where $m=0,\,\pm 1,\,\pm2,\dots,\,\pm\infty$ and $n=0,\,\pm 1,\,\pm2,\dots,\,\pm\infty$. 
}\label{fig:SM-hexagonal}
\end{figure}

We next consider a hexagonal array of soft magnetic strips, in which soft magnetic strips of width $2w$ are regularly arranged with a unit cell of $2a\times 2b$ in the $xy$ plane, as shown in \fref{fig:SM-hexagonal}. 
The geometry of the hexagonal array of soft magnetic strips is exactly the same as that of the hexagonal array of superconducting strips shown in \fref{fig:SC-hexagonal}. 

The complex field, ${\cal H}(\zeta)$, and the complex potential ${\cal G}(\zeta)=\int_{ib}^{\zeta}{\cal H}(\zeta')\rmd\zeta'$ for the hexagonal array of soft magnetic strips based on the ideal soft magnet model are given by 
\begin{eqnarray}
	{\cal H}(\zeta) &= H_{0y} 
		+\rmi\, H_{0x}\frac{\eta_h(\zeta)\sqrt{\eta_h(\zeta)^2-k_h^{-2}}}{%
		\sqrt{\eta_h(\zeta)^2-\gamma_h^2}\sqrt{\eta_h(\zeta)^2-\beta_h^2}} , 
\label{eq:H(z)_hexagonal_SM}\\
	{\cal G}(\zeta) &= H_{0y}\zeta 
		+\frac{\rmi\,H_{0x}c_h}{k_h\sqrt{\beta_h^2-\gamma_h^2}}\, 
		F\left(\arcsin\sqrt{\frac{\beta_h^2-\gamma_h^2}{\eta_h(\zeta)^2-\gamma_h^2}}, 
		\kappa_h \right) , 
\label{eq:G(z)_hexagonal_SM}
\end{eqnarray}
where $\eta_h$, $k_h$, $c_h$, $\gamma_h$, $\beta_h$, and $\kappa_h$ are defined by \eref{eq:eta-hexagonal}, \eref{eq:b/a-K_hexagonal}, \eref{eq:c-hexagonal}, \eref{eq:gamma_h}, \eref{eq:beta_h}, and \eref{eq:kappa_h}, respectively. 
\Eref{eq:H(z)_hexagonal_SM} and \eref{eq:G(z)_hexagonal_SM} are obtained simply by exchanging $H_{0y}\leftrightarrow\rmi H_{0x}$ in \eref{eq:H(z)_hexagonal} and \eref{eq:G(z)_hexagonal}, respectively. 

When the hexagonal array of soft magnetic strips is exposed to a {\em perpendicular} magnetic field along the $y$ axis, \eref{eq:H(z)_hexagonal_SM} shows that ${\cal H}(\zeta)=H_{0y}$ for $H_{0x}=0\neq H_{0y}$. 
The macroscopic permeability for a perpendicular field is equal to the vacuum permeability, $\mu_{\perp}/\mu_0=1$, for the thin-strip limit.  

\begin{figure}[bthp]
\center\includegraphics{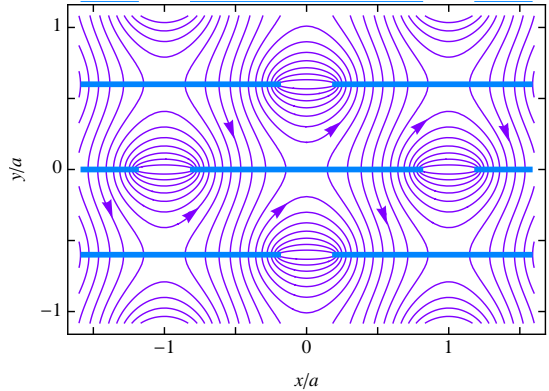}
\caption{%
Magnetic field lines in the hexagonal array of soft magnetic strips (shown as solid horizontal bars) exposed to a parallel magnetic field for $w/a=0.8$ and $b/a=0.3$. 
}\label{fig:field-lines_SM-hexa}
\end{figure}

In contrast, when the hexagonal array of soft magnetic strips is exposed to a {\em parallel} magnetic field along the $x$ axis ($H_{0x}\neq 0$ and $H_{0y}=0$), the magnetic field is disturbed by the soft magnetic strips. 
\Fref{fig:field-lines_SM-hexa} shows the magnetic field lines as the contour lines of $\mbox{Re}\,{\cal G}(x+\rmi y)$ obtained from \eref{eq:G(z)_hexagonal_SM}. 

The definitions of the macroscopic magnetic field, $\mean{H_x}$, and the magnetization, $\mean{M_x}$, for the hexagonal array are the same as those used for the rectangular array, and are given by \eref{eq:<Hx>_rectangular} and \eref{eq:Mx}, respectively. 
However, the definition of the macroscopic magnetic induction, $\mean{B_x}$, for the hexagonal array given by \eref{eq:<Bx>_rectangular} is inconsistent with the macroscopic relationship given by \eref{eq:Bx-Hx-Mx}. 
Therefore, we use the modified definition of $\mean{B_x}$ for the hexagonal array, 
\begin{equation}
	\frac{\mean{B_x}}{\mu_0}\equiv \frac{1}{2b}\left[\int_0^{2b}H_x(a,y)\rmd y 
		+\int_0^a H_y(x,2b-\epsilon)\rmd x\right] . 
\label{eq:<Bx>_hexagonal}
\end{equation}
For the hexagonal array, the macroscopic quantities defined by \eref{eq:<Hx>_rectangular}, \eref{eq:Mx}, and \eref{eq:<Bx>_hexagonal} satisfy \eref{eq:Bx-Hx-Mx}, as shown in \ref{app:sm-hexa}. 

The macroscopic parallel permeability, $\mu_{\rm \parallel sm,h}= \mean{B_x}/\mean{H_x}$, for the hexagonal array of soft magnetic strips is obtained from \eref{eq:<Hx>_rectangular}, \eref{eq:H(z)_hexagonal_SM}, and \eref{eq:<Bx>_hexagonal}, as 
\begin{equation}
	\frac{\mu_{\parallel\rm sm,h}}{\mu_0}= 
		\frac{a}{2b}\frac{{\bf K}(\sqrt{1-\kappa_h^2})}{{\bf K}(\kappa_h)} , 
\label{eq:mu-para_SM-hexa}
\end{equation}
where $\kappa_h$ is given by \eref{eq:kappa_h}. 
Note that $\mu_{\perp\rm sc,h}$ given by \eref{eq:mu-perp_SC-hexa} and $\mu_{\parallel\rm sm,h}$ given by \eref{eq:mu-para_SM-hexa} hold the simple relationship $\mu_{\parallel\rm sm,h}=\mu_0^2/\mu_{\perp\rm sc,h}$. 
\Fref{fig:mu-perp_SC-hexa} shows plots of $\mu_0/\mu_{\perp\rm sm,h}$ versus $w/a$ obtained from \eref{eq:b/a-K_hexagonal}, \eref{eq:c-hexagonal}, \eref{eq:kappa_h}, and \eref{eq:mu-perp_SC-hexa}. 
We can obtain a large parallel permeability, $\mu_{\parallel\rm sm,h}/\mu_0\gg 1$, for a wide range of $0.5<w/a<1$, when $b/a\ll 1$

\section{Hybrid arrays of superconducting and soft magnetic strips
\label{sec:SC-SM-array}}
We investigate the magnetic field distribution and macroscopic permeability of 2D arrays composed of both superconducting strips and soft magnetic strips. 
Here we consider the case when both superconducting strips and soft magnetic strips are parallel to the $xz$ plane~\cite{sc-sm-array}.

\subsection{Rectangular array of superconducting and soft magnetic strips}
\begin{figure}[bthp]
\center\includegraphics{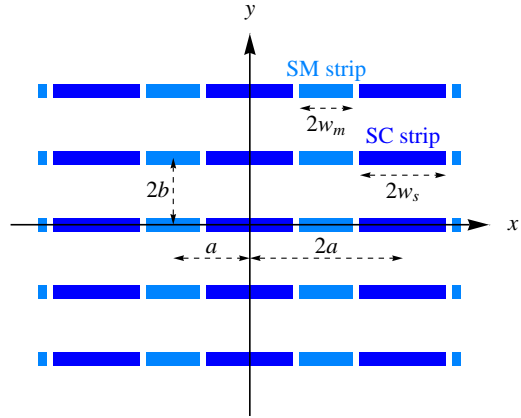}
\caption{%
Rectangular hybrid array of superconducting strips of width $2w_s$ and soft magnetic strips of width $2w_m$, where $w_s+w_m\leq a$. 
The solid horizontal bars show the cross sections of the superconducting (SC) strips and soft magnetic (SM) strips in the $xy$ plane. 
The hybrid array is a combination of the rectangular array of superconducting strips shown in \fref{fig:SC-rectangular} and the rectangular array of soft magnetic strips shown in \fref{fig:SM-rectangular}. 
The origin of the soft magnetic strip array is shifted from $(x,y)=(0,0)$ to $(a,0)$. 
}\label{fig:SC-SM-rectangular}
\end{figure}

We consider the hybrid array shown in \fref{fig:SC-SM-rectangular}, which is composed of the rectangular array of superconducting strips shown in \fref{fig:SC-rectangular} and the rectangular array of soft magnetic strips shown in \fref{fig:SM-rectangular}. 

The complex field ${\cal H}(\zeta)$ for the hybrid rectangular array of superconducting and soft magnetic strips is given by 
\begin{equation}
	{\cal H}(\zeta)= H_{0y}\frac{\eta_r(\zeta)}{\sqrt{\eta_r(\zeta)^2-\gamma_{rs}^2}} 
		+\rmi\,H_{0x}\frac{\eta_r(\zeta-a)}{\sqrt{\eta_r(\zeta-a)^2-\gamma_{rm}^2}} , 
\label{eq:H(z)_rectangular_sc-sm}
\end{equation}
where $\gamma_{rs}= \eta_r(w_s)= \mbox{sn}(w_s/c_r,k_r)$, $\gamma_{rm}= \eta_r(w_m-a)= \mbox{cn}(w_m/c_r,k_r)/\mbox{dn}(w_m/c_r,k_r)$, and $\eta_r(\zeta)$ is given by \eref{eq:eta-rectangular}. 
\Eref{eq:H(z)_rectangular_sc-sm} corresponds to the combination of \eref{eq:H(z)_rectangular} and \eref{eq:H(z)_rectangular_SM}. 
In a perpendicular magnetic field, $H_{0x}=0\neq H_{0y}$, the field distribution is determined by the arrangement of superconducting strips, and is not affected by the thin soft magnetic strips. 
In a parallel magnetic field, $H_{0y}=0\neq H_{0x}$, on the other hand, the field distribution is determined by the arrangement of soft magnetic strips, and is not affected by the thin superconducting strips. 

\begin{figure}[bthp]
\center\includegraphics{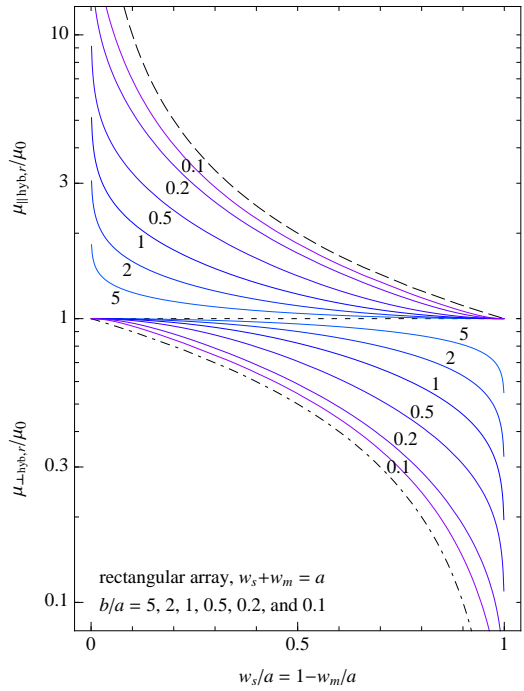}
\caption{%
Effective permeability for a perpendicular field, $\mu_{\perp\rm hyb,r}$, and for a parallel field, $\mu_{\parallel\rm hyb,r}$, of the rectangular hybrid array of superconducting strips and soft magnetic strips as a function of $w_s/a=1-w_m/a$ for $b/a=5,\,2,\,1,\,0.5,\,0.2$, and $0.1$. 
Dotted line corresponds to $\mu_{\perp}/\mu_0=\mu_{\parallel}/\mu_0=1$ for $b/a\to \infty$, the chained line to $\mu_{\perp}/\mu_0=1-w_s/a$ for $b/a\to 0$, and the dashed line to $\mu_{\parallel}/\mu_0=(1-w_m/a)^{-1}$ for $b/a\to 0$. 
}\label{fig:mu_SC-SM-rect}
\end{figure}

The resulting macroscopic permeability for a perpendicular field $\mu_{\perp\rm hyb,r}$ and that for a parallel field $\mu_{\parallel\rm hyb,r}$ are respectively given by 
\begin{eqnarray}
	\frac{\mu_{\perp\rm hyb,r}}{\mu_0} &= 
		\frac{b}{a}\frac{{\bf K}(\kappa_{rs})}{{\bf K}(\sqrt{1-\kappa_{rs}^2})} ,
\label{eq:mu-perp_SC-SM-rect}\\
	\frac{\mu_{\parallel\rm hyb,r}}{\mu_0} &= 
		\frac{a}{b}\frac{{\bf K}(\sqrt{1-\kappa_{rm}^2})}{{\bf K}(\kappa_{rm})} , 
\label{eq:mu-para_SC-SM-rect}
\end{eqnarray}
where $\kappa_{rs}=k_r\mbox{cn}(w_s/c_r,k_r)/\mbox{dn}(w_s/c_r,k_r)$ and $\kappa_{rm}=k_r\mbox{cn}(w_m/c_r,k_r)/\mbox{dn}(w_m/c_r,k_r)$. 
For small stack periodicity, $b/a\ll 1$, \eref{eq:mu-perp_SC-SM-rect} and \eref{eq:mu-para_SC-SM-rect} reduce to 
\begin{eqnarray}
	\frac{\mu_{\perp\rm hyb,r}}{\mu_0} &\simeq 
		1-\frac{w_s}{a}+\frac{2b}{\pi a}\ln 2 ,
\label{eq:mu-perp_SC-SM-rect_small-spacings}\\
	\frac{\mu_{\parallel\rm hyb,r}}{\mu_0} &\simeq 
		\left(1-\frac{w_m}{a}+\frac{2b}{\pi a}\ln 2\right)^{-1} .
\label{eq:mu-para_SC-SM-rect_small-spacings}
\end{eqnarray}
Equation \eref{eq:mu-perp_SC-SM-rect_small-spacings} is not accurate near $w_s/a\simeq 0$ or $1$, and \eref{eq:mu-para_SC-SM-rect_small-spacings} is not accurate near $w_m/a\simeq 0$ or $1$. 
If $w_s=w_m=a/2$ and $b/a\ll 1$, then $\mu_{\perp\rm hyb,r}/\mu_0\simeq \mu_0/\mu_{\parallel\rm hyb,r}\simeq 1/2$. 
\Fref{fig:mu_SC-SM-rect} shows plots of $\mu_{\perp\rm hyb,r}/\mu_0$ and $\mu_{\parallel\rm hyb,r}/\mu_0$ versus $w_s/a=1-w_m/a$ for the case where $w_s+w_m=a$, calculated from \eref{eq:mu-perp_SC-SM-rect} and \eref{eq:mu-para_SC-SM-rect}. 

\subsection{Hexagonal array of superconducting and soft magnetic strips%
\label{sec:hybrid-hexagonal}}
\begin{figure}[bthp]
\center\includegraphics{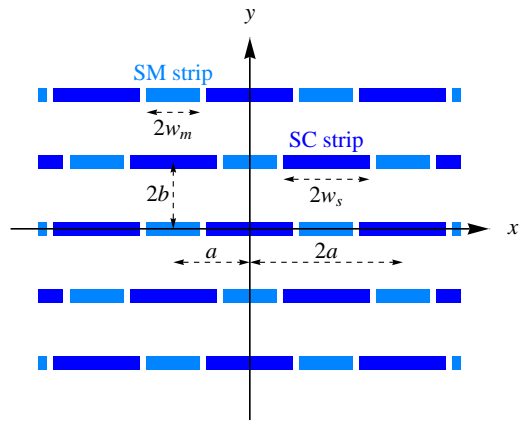}
\caption{%
Hexagonal hybrid array of superconducting strips of width $2w_s$ and soft magnetic strips of width $2w_m$, where $w_s+w_m\leq a$. 
The solid horizontal bars show the cross section of the superconducting strips and the soft magnetic strips in the $xy$ plane. 
The hybrid array is a combination of the hexagonal array of superconducting strips shown in \fref{fig:SC-hexagonal} and the hexagonal array of soft magnetic strips shown in \fref{fig:SM-hexagonal}. The origin of the magnetic strip array is shifted from $(x,y)=(0,0)$ to $(a,0)$. 
}\label{fig:SC-SM-hexagonal}
\end{figure}

We next consider the hybrid array shown in \fref{fig:SC-SM-hexagonal}, which is composed of the hexagonal array of superconducting strips shown in \fref{fig:SC-hexagonal} and the hexagonal array of soft magnetic strips shown in \fref{fig:SM-hexagonal}. 

The complex field ${\cal H}(\zeta)$ for the hybrid hexagonal array of superconducting and soft magnetic strips is given by 
\begin{eqnarray}
	{\cal H}(\zeta) =& H_{0y}\frac{\eta_h(\zeta)\sqrt{\eta_h(\zeta)^2-k_h^{-2}}}{%
		\sqrt{\eta_h(\zeta)^2-\gamma_{hs}^2}\sqrt{\eta_h(\zeta)^2-\beta_{hs}^2}} 
\nonumber\\
		& +\rmi\, H_{0x}\frac{\eta_h(\zeta-a)\sqrt{\eta_h(\zeta-a)^2-k_h^{-2}}}{%
		\sqrt{\eta_h(\zeta-a)^2-\gamma_{hm}^2}\sqrt{\eta_h(\zeta-a)^2-\beta_{hm}^2}} , 
\label{eq:H(z)_hexagonal_sc-sm}
\end{eqnarray}
where $\gamma_{hs}= \eta_h(w_s)= \mbox{sn}(w_s/c_h,k_h)$, $\gamma_{hm}= \eta_h(w_m-a)= \mbox{cn}(w_m/c_h,k_h)/\mbox{dn}(w_m/c_h,k_h)$, and $\eta_h(\zeta)$ is given by \eref{eq:eta-hexagonal}. 
\Eref{eq:H(z)_hexagonal_sc-sm} corresponds to the combination of \eref{eq:H(z)_hexagonal} and \eref{eq:H(z)_hexagonal_SM}. 
In a perpendicular magnetic field, $H_{0x}=0\neq H_{0y}$, the field distribution is determined by the arrangement of superconducting strips, and is not affected by the thin soft magnetic strips. 
In a parallel magnetic field, $H_{0y}=0\neq H_{0x}$, on the other hand, the field distribution is determined by the arrangement of soft magnetic strips, and is not affected by the thin superconducting strips. 

\begin{figure}[bthp]
\center\includegraphics{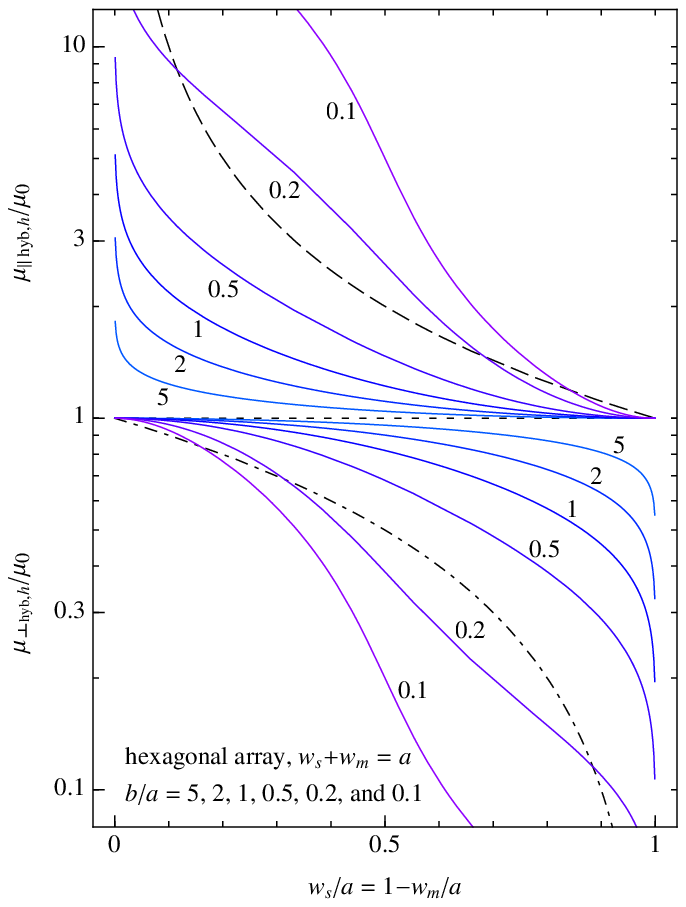}
\caption{%
Effective permeability for a perpendicular field, $\mu_{\perp\rm hyb,h}$, and for a parallel field, $\mu_{\parallel\rm hyb,h}$, of a hexagonal hybrid array of superconducting strips and soft magnetic strips as a function of $w_s/a=1-w_m/a$ for $b/a=5,\,2,\,1,\,0.5,\,0.2$, and $0.1$. 
Dotted line corresponds to $\mu_{\perp}/\mu_0=\mu_{\parallel}/\mu_0=1$ for $b/a\to \infty$. Chained line, which corresponds to $\mu_{\perp}/\mu_0=1-w_s/a$, and dashed line, which corresponds to $\mu_{\parallel}/\mu_0=(1-w_m/a)^{-1}$, are shown for comparison with \fref{fig:mu_SC-SM-rect}. 
}\label{fig:mu_SC-SM-hexa}
\end{figure}

The resulting macroscopic permeability for a perpendicular field, $\mu_{\perp}$, and that for a parallel field, $\mu_{\parallel}$, are given by 
\begin{eqnarray}
	\frac{\mu_{\perp\rm hyb,h}}{\mu_0} &= 
		\frac{2b}{a}\frac{{\bf K}(\kappa_{hs})}{{\bf K}(\sqrt{1-\kappa_{hs}^2})} ,
\label{eq:mu-perp_SC-SM-hexa}\\
	\frac{\mu_{\parallel\rm hyb,h}}{\mu_0} &= 
		\frac{a}{2b}\frac{{\bf K}(\sqrt{1-\kappa_{hm}^2})}{{\bf K}(\kappa_{hm})} , 
\label{eq:mu-para_SC-SM-hexa}
\end{eqnarray}
respectively, where 
\begin{eqnarray}
	\kappa_{hs} &= \left[1+\frac{k_h^{-2}-1}{\mbox{cn}^4(w_s/c_h,k_h)}\right]^{-1/2} , 
\label{eq:kappa_hs}\\
	\kappa_{hm} &= \left[1+\frac{k_h^{-2}-1}{\mbox{cn}^4(w_m/c_h,k_h)}\right]^{-1/2} , 
\label{eq:kappa_hm}
\end{eqnarray}
For a small stack periodicity, $b/a\ll 1$, \eref{eq:mu-perp_SC-SM-hexa} and \eref{eq:mu-para_SC-SM-hexa} reduce to 
\begin{equation}
	\frac{\mu_{\perp\rm hyb,h}}{\mu_0}\simeq \cases{%
		1-\frac{2w_s}{a} +\frac{8b}{\pi a}\ln2 
		& for $\displaystyle 0<w_s/a<1/2$ \\
		\left(\frac{2b}{a}\right)^2 \left(\frac{2w_s}{a}-1 +\frac{8b}{\pi a}\ln2\right)^{-1} 
		& for $1/2<w_s/a<1$ } , 
\label{eq:mu-perp_SC-SM-hexa_small-spacing}
\end{equation}
\begin{equation}
	\frac{\mu_{\parallel\rm hyb,h}}{\mu_0}\simeq \cases{%
		\left(1-\frac{2w_m}{a} +\frac{8b}{\pi a}\ln2\right)^{-1} 
		& for $\displaystyle 0<w_m/a<1/2$ \\
		\left(\frac{a}{2b}\right)^2 \left(\frac{2w_m}{a}-1 +\frac{8b}{\pi a}\ln2\right) 
		& for $1/2<w_m/a<1$ } . 
\label{eq:mu-parallel_SC-SM-hexa_small-spacing}
\end{equation}
Equation \eref{eq:mu-perp_SC-SM-hexa_small-spacing} is not accurate near $w_s/a\simeq 0$, $1/2$, or $1$, and \eref{eq:mu-parallel_SC-SM-hexa_small-spacing} is not accurate near $w_m/a\simeq 0$, $1/2$, or $1$. 
If $w_s=w_m=a/2$ and $b/a\ll 1$, then $\mu_{\perp\rm hyb,h}/\mu_0\simeq \mu_0/\mu_{\parallel\rm hyb,h}\simeq 2b/a\ll 1$. 
\Fref{fig:mu_SC-SM-hexa} shows plots of $\mu_{\perp\rm hyb,h}/\mu_0$ and $\mu_{\parallel\rm hyb,h}/\mu_0$ versus $w_s/a=1-w_m/a$ for the case where $w_s+w_m=a$, calculated from \eref{eq:mu-perp_SC-SM-hexa} and \eref{eq:mu-para_SC-SM-hexa}. 

\section{Discussion and summary
\label{sec:summmary}}
One of the most interesting applications of dc magnetic metamaterials is magnetic cloaking. 
We explore the possibility of the dc magnetic cloaking with a cylindrical tube of the magnetic metamaterial occupying the region $R_1<\rho<R_2$, where $R_1$ and $R_2$ are the inner and outer radii, respectively, and $(\rho,\theta,z)$ denotes the cylindrical coordinates. 
When the metamaterial tube is exposed to a transverse magnetic field, which is perpendicular to the $z$ axis, the magnetic field inside the metamaterial tube ($0<\rho<R_1$) should be zero, whereas the magnetic field outside the tube ($\rho>R_2$) should be undisturbed. 
This cylindrical cloaking can be achieved, when the radial and azimuthal permeabilities are respectively given by~\cite{Cummer06,Schurig06,Yaghjian08,Sanchez11} 
\begin{equation}
	\mu_{\rho}/\mu_0=1-R_1/\rho , \quad
	\mu_{\theta}/\mu_0=(1-R_1/\rho)^{-1} . 
\label{eq:mu_r-theta_cloaking}
\end{equation}
Therefore, anisotropic permeabilities where $0<\mu_{\rho}/\mu_0<1<\mu_{\theta}/\mu_0<\infty$ and $\mu_{\rho}/\mu_0=\mu_0/\mu_{\theta}$ are required; the hybrid hexagonal array of superconducting and soft magnetic strips investigated in Sec. \ref{sec:hybrid-hexagonal} may achieve this. 
If superconducting strips and soft magnetic strips are arranged such that their wide surfaces are perpendicular to the radial direction of the cylindrical metamaterial tube, the permeabilities should follow $\mu_{\perp}=\mu_{\rho}$ and $\mu_{\parallel}=\mu_{\theta}$.  
Equations \eref{eq:mu-perp_SC-SM-hexa} and \eref{eq:mu-para_SC-SM-hexa} show that $0<\mu_{\perp}/\mu_0<1<\mu_{\parallel}/\mu_0<\infty$ and $\mu_{\perp}/\mu_0=\mu_0/\mu_{\parallel}$, for superconducting and soft magnetic strips of identical widths, $w_s=w_m$. 
By adjusting the width, $w_s=w_m$, and the array periodicity, $a$, as a function of $\rho$, \eref{eq:mu_r-theta_cloaking} can be satisfied approximately. 
However, the magnetic cloaking would be incomplete, because of the $\mu_{\perp}\to 0$ and $\mu_{\parallel}\to\infty$ singularities at $\rho\to R_1$ in \eref{eq:mu_r-theta_cloaking}. 
Other types of magnetic cloaking devices composed of superconductor-magnet bilayers which avoid these singular permeabilities have also been proposed and experimentally verified~\cite{Sanchez11,Narayana11,Gomory12}. 

We have theoretically investigated the field distribution in infinite 2D arrays of thin superconducting and soft magnetic strips, which are essential structures for dc magnetic metamaterials. 
The geometry of the thin strips produced anisotropy in the macroscopic permeability, $\mu_{\perp}$, when the applied magnetic field was perpendicular to the wide surface of the strips and in $\mu_{\parallel}$ when it was parallel. 
The macroscopic permeability of the 2D arrays of superconducting strips showed that $0<\mu_{\perp}/\mu_0\ll \mu_{\parallel}/\mu_0\simeq 1$. 
The behavior of the soft magnetic strips was analogously dual to that of the superconducting strips, and the macroscopic permeability of the 2D arrays of the soft magnetic strips showed that $1\simeq \mu_{\perp}/\mu_0\ll \mu_{\parallel}/\mu_0$. 
Hybrid arrays of the superconducting and soft magnetic strips exhibited strongly anisotropic macroscopic permeability, $0<\mu_{\perp}/\mu_0\ll 1\ll \mu_{\perp}/\mu_0$. 
We have also investigated two array configurations, and showed that the hexagonal arrays were better for producing strongly anisotropic permeability than the rectangular arrays. 

We adopted simple models for superconductors and soft magnets; the magnetic field was completely shielded in the superconductors, and the soft magnets had an infinite permeability, zero hysteresis, and an infinite saturation field. 
More realistic models of superconductors and soft magnets could be investigated by numerical simulations~\cite{Navau09,Mawatari12,Gomory10}. 
We focused on two-dimensional arrays of strips that have infinite length $L_z\to\infty$ along the $z$ axis. 
Three-dimensional rectangular arrays of superconducting square plates (i.e., $L_z=2w$ in our notation) was numerically investigated by Navau {\em et al.}~\cite{Navau09}, who showed that the lower limit of the macroscopic permeability is $\mu_{\perp}/\mu_0=1-(w/a)^2$, in contrast to the lower limit $\mu_{\rm\perp sc,r}/\mu_0=1-w/a$ for the two-dimensional rectangular array shown as the chained line in \fref{fig:mu-perp_SC-rect}. 
Numerical simulation for such realistic three-dimensional arrays of superconducting and soft magnetic plates should also be investigated as future works. 
Furthermore, the details of magnetic metamaterial design should be investigated for magnetic cloaking and other possible applications.

\appendix
\section{Macroscopic relationship between $\mean{\bi B}$, $\mean{\bi H}$, and $\mean{\bi M}$
\label{sec:B-H-M}}
In this appendix we examine the definition of the macroscopic magnetic induction $\mean{\bi B}$ and that of the macroscopic magnetic field $\mean{\bi H}$ to be consistent with the macroscopic relationship between $\mean{\bi B}$, $\mean{\bi H}$, and the magnetization $\mean{\bi M}$. 

\subsection{Rectangular array of superconducting strips
\label{app:sc-rect}}
Because the current density in superconducting strips is given by $j_z=\partial H_y/\partial x-\partial H_x/\partial y$, the magnetization of superconducting strips $\mean{M_y}$ defined by \eref{eq:My} is calculated as 
\begin{eqnarray}
	4ab\mean{M_y} 
	&= -\int^{+a}_{-a}\rmd x\int^{2b-\epsilon}_{-\epsilon}\rmd y \,
		x\left(\frac{\partial H_y}{\partial x} -\frac{\partial H_x}{\partial y}\right) 
\nonumber\\
	&= -\int^{2b}_{0}\rmd y \int^{+a}_{-a} x\frac{\partial H_y}{\partial x} \rmd x 
		+\int^{+a}_{-a}x\,\rmd x \int^{2b-\epsilon}_{-\epsilon} 
			\frac{\partial H_x}{\partial y} \rmd y
\nonumber\\
	&= -\int^{2b}_{0}\rmd y \left[2a H_y(a,y) -\int^{+a}_{-a}H_y(x,y)\rmd x \right] 
\nonumber\\
	&\quad 	+\int^{+a}_{-a}x\,\rmd x \left[H_x(x,2b-\epsilon)-H_x(x,-\epsilon)\right] , 
\label{eq:My_sc-rect_cal}
\end{eqnarray}
where we used $H_y(-a,y)=H_y(a,y)$. 
For the rectangular array of superconducting strips, substitution of $H_x(x,2b-\epsilon)=H_x(x,-\epsilon)$ into \eref{eq:My_sc-rect_cal} yields 
\begin{equation}
	\mean{M_y}= -\frac{1}{2b}\int^{2b}_{0} H_y(a,y)\rmd y 
		+\frac{1}{4ab}\int^{2b}_{0} \rmd y \int^{+a}_{-a}H_y(x,y)\rmd x . 
\label{eq:My_sc-rect_cal2}
\end{equation}
Using \eref{eq:<By>_rectangular} and \eref{eq:<Hy>_rectangular}, we verify that \eref{eq:My_sc-rect_cal2} corresponds to \eref{eq:By-Hy-My}. 
In other words, the definitions of \eref{eq:<By>_rectangular} and \eref{eq:<Hy>_rectangular} are consistent with \eref{eq:By-Hy-My}.

\subsection{Hexagonal array of superconducting strips
\label{app:sc-hexa}}
For the hexagonal array of superconducting strips, the boundary condition of $H_x(x,-\epsilon)=-H_x(a-x,2b-\epsilon)$ leads to  
\begin{eqnarray}
	\lefteqn{ \int^{+a}_{-a} x \left[H_x(x,2b-\epsilon)-H_x(x,-\epsilon)\right] \rmd x } 
\nonumber\\
	&= 2 \int^{a}_{0} x \left[H_x(x,2b-\epsilon) +H_x(a-x,2b-\epsilon)\right] \rmd x
\nonumber\\
	&= 2a \int^a_0 H_x(x,2b-\epsilon) \rmd x . 
\label{eq:Hx(x,2b)_sc-hexa}
\end{eqnarray}
\Eref{eq:My_sc-rect_cal} is also valid for the hexagonal array of superconducting strip, and substitution of \eref{eq:Hx(x,2b)_sc-hexa} into \eref{eq:My_sc-rect_cal} yields 
\begin{eqnarray}
	\mean{M_y} &= {}-\frac{1}{2b}\left[ \int^{2b}_{0} H_y(a,y)\rmd y 
		-\int^{a}_{0}H_x(x,2b-\epsilon) \rmd x \right] 
\nonumber\\
	&\quad +\frac{1}{4ab}\int^{2b}_{0} \rmd y \int^{+a}_{-a}H_y(x,y)\rmd x . 
\label{eq:My_sc-hexa_cal2}
\end{eqnarray}
Using \eref{eq:<By>_rectangular} and \eref{eq:<Hy>_hexagonal}, we verify that \eref{eq:My_sc-hexa_cal2} corresponds to \eref{eq:By-Hy-My}. 
In other words, the definitions of \eref{eq:<By>_rectangular} and \eref{eq:<Hy>_hexagonal} are consistent with \eref{eq:By-Hy-My}.

\subsection{Rectangular array of soft magnetic strips
\label{app:sm-rect}}
Because the effective magnetic charge density in soft magnetic strips is given by $\rho_m=\mu_0(\partial H_x/\partial x+\partial H_y/\partial y)$, the magnetization of soft magnetic strips $\mean{M_x}$ defined by \eref{eq:Mx} is calculated as 
\begin{eqnarray}
	4ab\mean{M_x} 
	&= \int^{+a}_{-a}\rmd x\int^{2b-\epsilon}_{-\epsilon}\rmd y \,
		x\left(\frac{\partial H_x}{\partial x} +\frac{\partial H_y}{\partial y}\right) 
\nonumber\\
	&= \int^{2b}_{0}\rmd y \int^{+a}_{-a} x\frac{\partial H_x}{\partial x} \rmd x 
		+\int^{+a}_{-a}x\,\rmd x \int^{2b-\epsilon}_{-\epsilon} 
			\frac{\partial H_y}{\partial y} \rmd y
\nonumber\\
	&= \int^{2b}_{0}\rmd y \left[ 2a H_x(a,y) -\int^{+a}_{-a}H_x(x,y)\rmd x \right] 
\nonumber\\
	&\quad 	+\int^{+a}_{-a}x\,\rmd x \left[H_y(x,2b-\epsilon)-H_y(x,-\epsilon)\right] , 
\label{eq:Mx_sm-rect_cal}
\end{eqnarray}
where we used $H_x(-a,y)=H_x(a,y)$. 
For the rectangular array of soft magnetic strips, substitution of $H_y(x,2b-\epsilon)=H_y(x,-\epsilon)$ into \eref{eq:Mx_sm-rect_cal} yields 
\begin{equation}
	\mean{M_x}= \frac{1}{2b}\int^{2b}_{0} H_x(a,y)\rmd y 
		-\frac{1}{4ab}\int^{2b}_{0} \rmd y \int^{+a}_{-a}H_x(x,y)\rmd x . 
\label{eq:Mx_sm-rect_cal2}
\end{equation}
Using \eref{eq:<Bx>_rectangular} and \eref{eq:<Hx>_rectangular}, we verify that \eref{eq:Mx_sm-rect_cal2} corresponds to \eref{eq:Bx-Hx-Mx}. 
In other words, the definitions of \eref{eq:<Bx>_rectangular} and \eref{eq:<Hx>_rectangular} are consistent with \eref{eq:Bx-Hx-Mx}.

\subsection{Hexagonal array of soft magnetic strips
\label{app:sm-hexa}}
For the hexagonal array of soft magnetic strips, the boundary condition of $H_y(x,-\epsilon)=-H_y(a-x,2b-\epsilon)$ leads to 
\begin{eqnarray}
	\lefteqn{ \int^{+a}_{-a} x \left[H_y(x,2b-\epsilon)-H_y(x,-\epsilon)\right] \rmd x } 
\nonumber\\
	&= 2 \int^{a}_{0} x \left[H_y(x,2b-\epsilon) +H_y(a-x,2b-\epsilon)\right] \rmd x
\nonumber\\
	&= 2a \int^a_0 H_y(x,2b-\epsilon) \rmd x . 
\label{eq:Hy(x,2b)_sm-hexa}
\end{eqnarray}
\Eref{eq:Mx_sm-rect_cal} is also valid for the hexagonal array of soft magnetic strip, and substitution of \eref{eq:Hy(x,2b)_sm-hexa} into \eref{eq:Mx_sm-rect_cal} yields 
\begin{eqnarray}
	\mean{M_x} &= \frac{1}{2b}\left[ \int^{2b}_{0} H_x(a,y)\rmd y 
		+\int^{a}_{0}H_y(x,2b-\epsilon) \rmd x \right] 
\nonumber\\
	&\quad -\frac{1}{4ab}\int^{2b}_{0} \rmd y \int^{+a}_{-a}H_x(x,y)\rmd x . 
\label{eq:My_sm-hexa_cal2}
\end{eqnarray}
Using \eref{eq:<Hx>_rectangular} and \eref{eq:<Bx>_hexagonal}, we verify that \eref{eq:My_sm-hexa_cal2} corresponds to \eref{eq:Bx-Hx-Mx}. 
In other words, the definitions of \eref{eq:<Hx>_rectangular} and \eref{eq:<Bx>_hexagonal} are consistent with \eref{eq:Bx-Hx-Mx}.

\section*{References}


\begin{thebibliography}{99}
\bibitem{Wood07}
Wood B and Pendry J B 2007
Metamaterials at zero frequency
\JPCM {\bf 19} 076208
\bibitem{Magnus08}
Magnus F, Wood B, Moore J, Morrison K, Perkins G, Fyson J, Wiltshire M C K, Caplin D, Cohen L F and Pendry J B 2008
A d.c. magnetic metamaterial 
{\em Nature Mater.} {\bf 7} 295
\bibitem{Navau09}
Navau C, Chen D-X, Sanchez A and Del-Valle N 2009
Magnetic properties of a dc metamaterial consisting of parallel square superconducting thin plates 
{\em Appl. Phys. Lett.} {\bf 94} 242501

\bibitem{Cummer06}
Cummer S A, Popa B-I, Schurig D and Smith D R 2006 
Full-wave simulations of electromagnetic cloaking structures 
{\em Phys. Rev. E} {\bf 74} 036621
\bibitem{Schurig06}
Schurig D, Mock J J, Justice B J, Cummer S A, Pendry J B, Starr A F, and Smith D R 2006 
Metamaterial Electromagnetic Cloak at Microwave Frequencies 
{\em Science} {\bf 314} 977--980
\bibitem{Yaghjian08}
Yaghjian A D and Maci S 2008
Alternative derivation of electromagnetic cloaks and concentrators
\NJP {\bf 10} 115022
\bibitem{Sanchez11}
Sanchez A, Navau C, Prat-Camps J, and Chen D-X 2011 
Antimagnets: controlling magnetic fields with superconductor-metamaterial hybrids
\NJP {\bf 13} 093034
\bibitem{Narayana11}
Narayana S and Sato Y 2011
DC Magnetic Cloak
{\em Advanced Materials} {\bf 24} 71--74
\bibitem{Gomory12}
G\"{o}m\"{o}ry F, Solovyov M, \v{S}ouc J, Navau C, Prat-Camps J and Sanchez A 2012 
Experimental Realization of a Magnetic Cloak
{\em Science} {\bf 335} 1466--1468
\bibitem{Mawatari12}
Mawatari Y, Navau C and Sanchez A 2012
Two-dimensional arrays of superconducting strips as dc magnetic metamaterials
\PR{\em B} {\bf 85} 134524

\bibitem{Landau-Lifschitz}
Landau L D and Lifschitz E M 1963  
{\em Electrodynamics of Continuous Media}, 
Theoretical Physics (Pergamon, Oxford, 1963), Vol. 8
\bibitem{Beth66}
Beth R A 1966 
Complex representation and computation of two-dimensional magnetic fields
\JAP {\bf 37} 2568

\bibitem{Pendry99}
Pendry J B, Holden A J, Robbins D J and Stewart W J 1999 
Magnetism from conductors and enhanced nonlinear phenomena 
{\em IEEE Trans. Microwave Theory Tech.} {\bf 47} 2075
\bibitem{Smith06}
Smith D R and Pendry J B 2006 
Homogenization of metamaterials by field averaging 
{\em J. Opt. Soc. Am. B} {\bf 23} 391

\bibitem{Bean62}
Bean C P (1962). 
Magnetization of hard superconductors. 
\PRL, 8, 250-253.

\bibitem{Gradshtein-Ryzhik}
Gradshtein I S and Ryzhik I M 1994
{\it Table of Integrals, Series, and Products}, 5th ed., 
(Academic, New York)
%
\bibitem{H0-Ha}
For linear magnetic materials investigated in the present paper (i.e., superconducting strips in the complete shielding state or ideal soft magnetic strips), neither $H_{0x}$ nor $H_{0y}$ affects the effective permeability. 
If we consider the nonliner magnetic response (e.g., superconducting strips in the critical state), the $H_{0y}$ and $H_{0x}$ need to be determined as functions of $H_{ay}$ and $H_{ax}$. 
The relationship between the complex field $H_{0y}+\rmi H_{0x}$ at $(x,y)=(0,b)$ and the applied field $H_{ay}+\rmi H_{ax}$ may be determined by considering the total shape (e.g., the demagnetizaition factor) of magnetic metamaterials. 

\bibitem{Mawatari08}
Mawatari Y 2008 
Magnetic field distributions around superconducting strips on ferromagnetic substrates 
{\em Phys. Rev. B} {\bf 77} 104505

\bibitem{Jackson}
Jackson J D 1975 
{\it Classical Electrodynamics}, 2nd ed., 
(Wiley, New York)
%
\bibitem{Chen02}
Chen D-X, Prados C, Pardo E, Sanchez A and Hernando A 2002
Transverse demagnetizing factors of long rectangular bars: I. Analytical expressions for extreme values of susceptibility 
\JAP {\bf 91} 5254
%
\bibitem{sc-sm-array}
We do not investigate the case when soft magnetic strips are vertical to superconducting strips, because the anisotropy in the macroscopic permeability is weak for such vertical hybrid arrays. 
%
\bibitem{Gomory10}
G\"{o}m\"{o}ry F, Vojen\v{c}iak M, Pardo E, Solovyov M and \v{S}ouc J 2010 
AC losses in coated conductors 
\SUST {\bf 23} 034012

\end{thebibliography}
\end{document}